\documentclass[aps,prl,showpacs,superscriptaddress,twocolumn]{revtex4-1}
\usepackage{amsmath, amssymb, amsthm}
\usepackage{mathtools}
\usepackage{mathrsfs}
\usepackage{bm}
\usepackage{slashed}   
\usepackage{graphicx}
\usepackage{multirow}
\usepackage{tikz}
\usepackage[caption=false]{subfig}
\usepackage{relsize}	
\usepackage{array}
\usepackage{float}
\usepackage{color}
\usepackage{xcolor}
\usepackage{soul}
\usepackage{ulem}
 \usepackage{longtable}
\usepackage{hyperref}
\hypersetup{colorlinks=true,linkcolor=magenta,anchorcolor=magenta,citecolor=magenta, filecolor=magenta,urlcolor=magenta,bookmarksnumbered=true, pdfview=FitB
}

\def\dd{{\mathrm{d}}}

\mathchardef\-="2D

\colorlet{darkgreen}{green!60!black}
\colorlet{brightyellow}{yellow!75!red}
\colorlet{orange}{red!50!yellow}
\colorlet{darkblue}{blue!60!black}
\colorlet{darkred}{red!80!black}
\colorlet{greenblue}{green!50!blue}

\makeatletter

\newcommand{\Rmnum}[1]{\expandafter\@slowromancap\romannumeral #1@}
\makeatother

\begin{document}
\title{Light-front holography with chiral symmetry breaking}

\author{Yang Li}
\affiliation{Department of Modern Physics, University of Sciences and Technology of China, Hefei 230026, China}
\affiliation{School of Nuclear Science and Technology, University of Chinese Academy of Sciences, Beijing 100049, China}
\affiliation{Department of Physics and Astronomy, Iowa State University, Ames, IA 50011, USA}

\author{James P. Vary}
\affiliation{Department of Physics and Astronomy, Iowa State University, Ames, IA 50011, USA}

\date{\today}
\begin{abstract}

We present an analytically solvable 3D light-front Hamiltonian model for hadrons that extends light-front holography by including finite mass quarks and a longitudinal confinement term.  We propose that the model is suitable as an improved analytic approximation to QCD at a low resolution scale. We demonstrate that it preserves desired Lorentz symmetries and it produces improved agreement with the experimental mass spectroscopy and other properties of the light mesons. Importantly, the model also respects chiral symmetry and the Gell-Mann-Oakes-Renner relation.

\end{abstract}

\maketitle

The Schr\"odinger equation, one of the greatest triumphs of modern physics, provides the first approximation to the structure of atoms. Subsequent improvements using perturbation techniques in quantum electrodynamics (QED) have led to one of the most precise predictions in physics {known as the Lamb shift}. {For the investigation of hadrons and nuclei, quantum chromodynamics (QCD)} is the relevant theory but it is non-perturbative. Two dynamical features of QCD, confinement and chiral symmetry breaking, are expected to emerge at the scale of hadrons. To this date, despite major advancements in numerical calculations, our understanding of how QCD supports these non-perturbative features remains incomplete. In light of the extraordinary challenges in accurately describing the nonperturbative structure of hadrons, one is motivated to search for an analytic semiclassical approximation in QCD. 

In recent years, significant progress has been made in the Hamiltonian formulation of QCD quantized on the light front $x^+=t+z$. In particular, basis light-front quantization (BLFQ) provides a computational framework for solving light-front QCD (LFQCD) as relativistic quantum many-body theory \cite{Vary:2009gt}. The renormalization group procedure for effective particles (RGPEP) implements the Wilsonian view of QCD and provides a renormalized effective Hamiltonian that founded on first principles \cite{Glazek:2012qj}.  An immediate implication of RGPEP is the light-front Schr\"odinger equation, 
\begin{multline}\label{eqn:lfse}
\Big[ \frac{\vec k_\perp^2+m^2_q}{x} + \frac{\vec k_\perp^2+m^2_{\bar q}}{1-x} + V_\mathrm{eff} \Big] \psi_h (x, \vec k_\perp) \\= M^2_h \psi_h (x, \vec k_\perp).
\end{multline}
Here $x=p^+/P^+$ is the longitudinal momentum fraction of the quark, $\vec k_\perp = \vec p_\perp-x\vec P_\perp$ is the relative transverse momentum between the quark ($q$) and antiquark ($\bar q$). Hence the wave function $\psi_h(x, \vec k_\perp)$ is frame-independent and describes the relativistic intrinsic structure of the hadron ($h$). {$m_q$ ($m_{\bar q}$) is the effective quark (antiquark) mass.} $M_h$ is the mass of the hadron.  The effective potential $V_\mathrm{eff}$ plays a fundamental role in QCD, similar to the Coulomb potential in QED. An effective potential up to $O(\alpha_s)$ has been derived for heavy quarkonium from RGPEP \cite{Glazek:2017rwe}, signifying a first step along this path. 

In a dramatically different approach, light-front holography (LFH) determines the effective potential based on a unique mapping between the equation of motion of the string modes in the Anti-de Sitter (AdS) space \cite{Karch:2006pv} and the light-front Schr\"odinger equation of hadrons, as consistent with the conformal quantum mechanics and supersymmetry \cite{Brodsky:2013ar}. This semiclassical approximation is phenomenologically successful in hadron spectroscopy \cite{Brodsky:2006uqa, deTeramond:2014asa, Dosch:2015bca, Dosch:2016zdv}, including tetraquarks and exotica \cite{Nielsen:2018ytt, Zou:2019tpo}, form factors \cite{Brodsky:2006uqa, Brodsky:2008pf, Sufian:2016hwn}, and parton distributions\cite{deTeramond:2018ecg, Liu:2019vsn}. See Refs.~\cite{Brodsky:2014yha, Brodsky:2020ajy} for recent reviews.

A key step of LFH is the observation that the kinetic energy term (\ref{eqn:lfse}) in the chiral limit ($m_q=0$) depends only on a 
2D vector $\vec k_\perp/\sqrt{x(1-x)}$. Thus, to first approximation, the light-front Schr\"odinger equation (\ref{eqn:lfse}) reduces to 
a 2D equation in terms of $\vec \zeta_\perp = \sqrt{x(1-x)}\vec r_\perp$, where $\vec r_\perp$ 
is the transverse separation of the constituents:
\begin{equation}\label{eqn:lfse_perp}
\Big[ -\nabla^2_{\zeta_\perp} + V_\perp(\vec \zeta_\perp)\Big] \phi(\vec\zeta_\perp) = M^2_\perp \phi(\vec\zeta_\perp).
\end{equation}
This equation becomes identical to the equation of motion in the soft-wall AdS/QCD if $\zeta_\perp$ is identified with the fifth dimension $z$ in AdS space and the wave function $\phi(\zeta)$ is identified with the string modes $\Phi(z)$. This holographic mapping can be exactly verified using the matrix elements of the current operator and the energy-momentum tensor \cite{Brodsky:2006uqa, Brodsky:2008pf}. 
Thus the confining interaction is uniquely determined by the holographic boundary condition as $V_\perp (\zeta_\perp) = \kappa^4 \zeta_\perp^2 + 2\kappa(J-1)$, where $J$ is the total angular momentum. $\kappa$ is the strength of the holographic confinement. From Eq.~(\ref{eqn:lfse_perp}), the mass eigenvalues $M^2_{\perp} = 2\kappa^2 (2n+|m|+J)$, follow the Regge trajectories $M^2 \propto n,L,J$, where $n,m$ are the radial and angular quantum numbers in the transverse plane ($\vec \zeta_\perp$). $L$ is the orbital angular momentum ($m$ is its magnetic projection). The obtained wave functions $\phi_{nm}$ are 2D harmonic oscillator functions of the holographic variable $\zeta_\perp$ or its conjugate moment $k_\perp/\sqrt{x(1-x)}$. In particular, LFH predicts a massless pion ($n=m=J=0$) in the chiral limit, with the light-front wave function $\psi_\pi(x, \vec k_\perp) = (4\pi/\kappa)\exp\big[-\vec k_\perp^2/(2\kappa^2x(1-x))\big]$. 

To compare with physical mesons, quark masses need to be incorporated. Brodsky and de~Téramond adopted a natural ansatz, $k_\perp^2/x(1-x) \to (k_\perp^2+m_q^2)/x + (k_\perp^2+m_{\bar q}^2)/(1-x)$, known as the invariant mass ansatz (IMA) \cite{Brodsky:2008pf}. The resulting pion wave function reads,
$\psi_\pi(x, \vec k_\perp) = N(4\pi/\kappa)\exp\big[-(\vec k_\perp^2+m_{u,d}^2)/(2\kappa^2x(1-x))\big]$, 
where $N$ is a normalization constant, $m_{u,d} = (m_u + m_d)/2$ is the light quark mass. The pion mass becomes nonzero, 
\begin{equation}\label{eqn:Mpi_IMA}
M_\pi^2 = \int \frac{\dd x}{2x(1-x)} \frac{\dd^2k_\perp}{(2\pi)^3} \big| \psi_\pi(x, \vec k_\perp) \big|^2 \frac{m_q^2}{x(1-x)}.
\end{equation} 
In the vicinity of the chiral limit, the theory predicts a quadratic (up to a logarithm) quark mass dependence of the pion mass squared, $M_\pi^2 \approx 2m_{u,d}^2 (\ln \kappa^2/m_{u,d}^2-\gamma_\textsc{e})$, where $\gamma_\textsc{e}\approx 0.577216$ is the Euler's constant. 
On the other hand, in QCD, the pion is the Goldstone boson of chiral symmetry breaking. When finite quark masses are present, the pion mass is dictated by the partially conserved axial current in the form of the Gell-Mann-Oakes-Renner (GMOR) relation \cite{Gell-Mann:1968},
\begin{equation}  \label{eqn:GMOR}
 f^2_\pi M^2_\pi = 2 m_{u,d} \big|\langle \bar q q \rangle\big| + \mathcal O(m_{u,d}^2),
\end{equation}
where $\langle \bar q q \rangle$ is the vacuum quark condensate. $f_\pi$ is the pion decay constant.  
The GMOR relation predicts a linear quark mass dependence of the pion mass squared, $M^2_\pi \propto 2m_{u,d}$, in contrast to Eq.~(\ref{eqn:Mpi_IMA}). 

Is it possible to reconcile LFH and the chiral symmetry breaking? In this work, we show that incorporation of the longitudinal dynamics, a previously unexplored degree of freedom (d.o.f.) in LFH, produces both consistency with the chiral symmetry breaking as well as improvements in meson mass spectroscopy. A great advantage of this method is that the wave functions of states without longitudinal excitations, such as $\rho, \pi, K$, remain mostly intact. So does the 
phenomenological success of LFH on the properties of these states.

In general, we can assume the effective potential $V_\mathrm{eff}$ consists of a separable longitudinal confining term $V_\|$ in addition to the holographic confinement $V_\perp$, viz.~$V_\mathrm{eff} = V_\perp(\vec\zeta_\perp) + V_\|(x)$. Then, the light-front Schr\"odinger equation~(\ref{eqn:lfse}) implies the existence of a longitudinal equation,
\begin{equation}\label{eqn:lfse_long}
\Big[ \frac{m_q^2}{x} + \frac{m_{\bar q}^2}{1-x} + V_\| \Big] \chi(x) = M^2_\| \chi(x),
\end{equation}
where $\chi$ is the longitudinal wave function from the separation of variables of the full wave function $\psi(x, \vec k_\perp) = \phi(\vec k_\perp/\sqrt{x(1-x)})\chi(x)$, which is normalized as $\int\dd x |\chi(x)|^2 = 4\pi$. The full mass squared eigenvalues $M^2 = M_\perp^2+M^2_\|$. 

\begin{figure*}
\centering 
\includegraphics[width=0.45\textwidth]{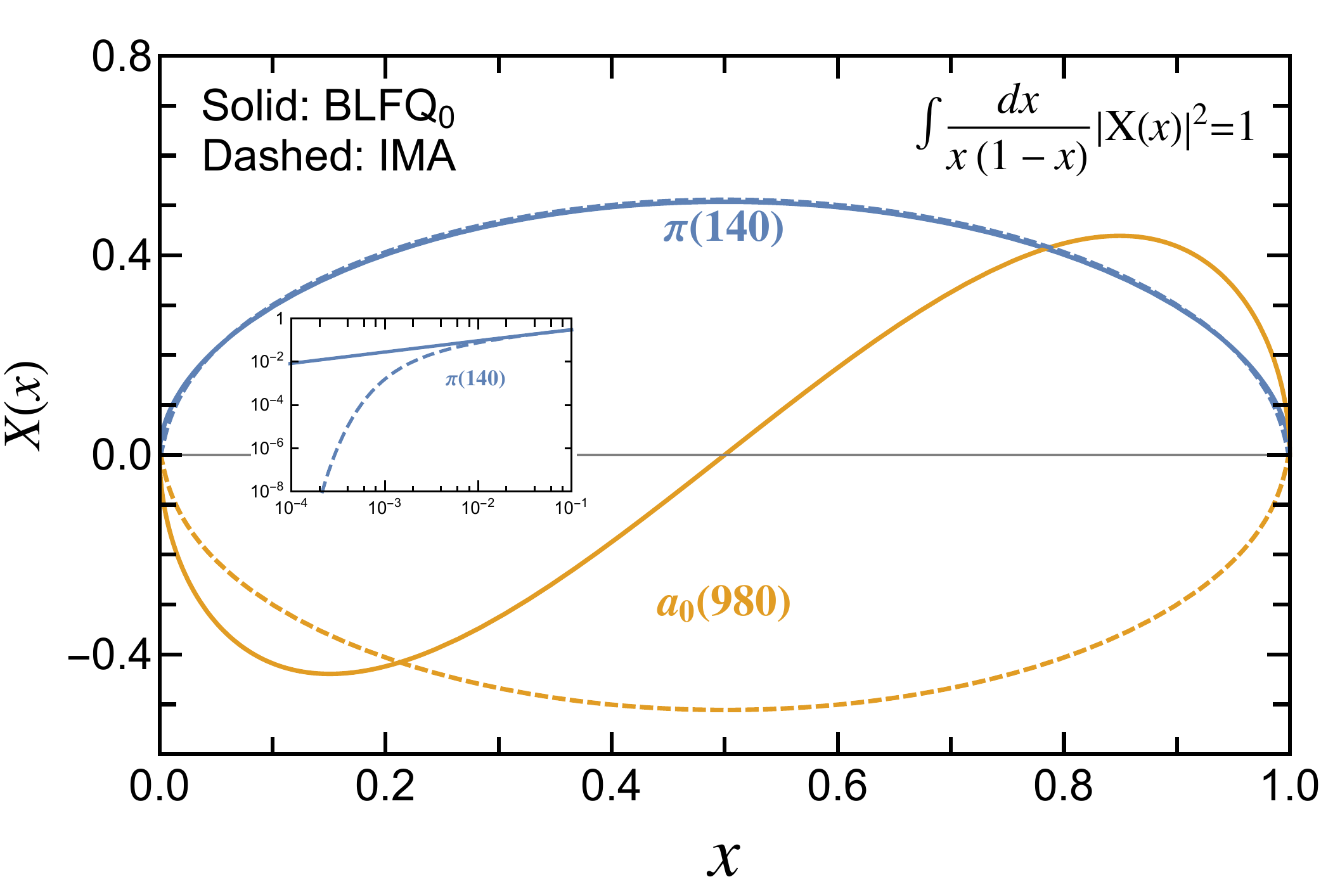}
\includegraphics[width=0.45\textwidth]{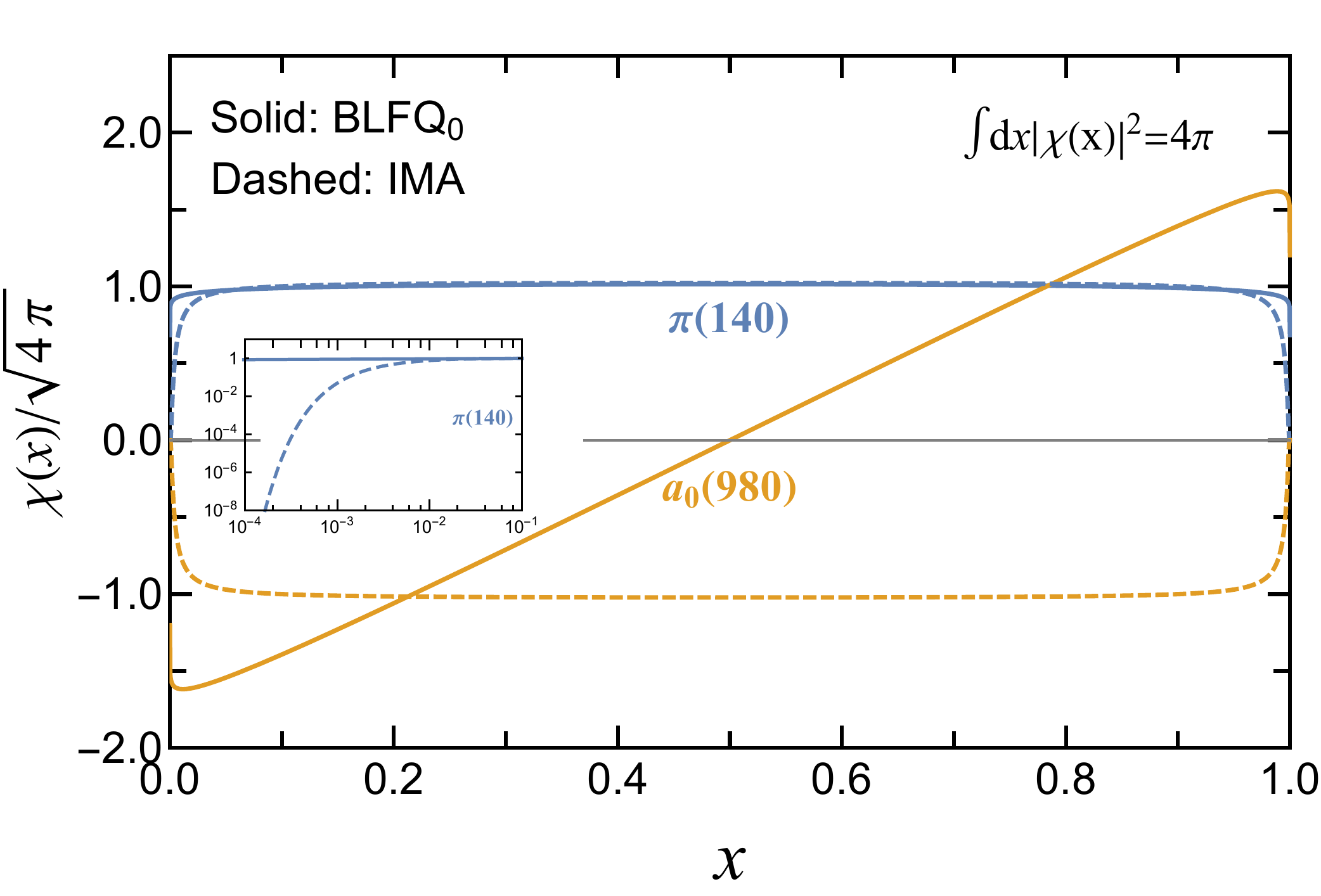}
\includegraphics[width=0.45\textwidth]{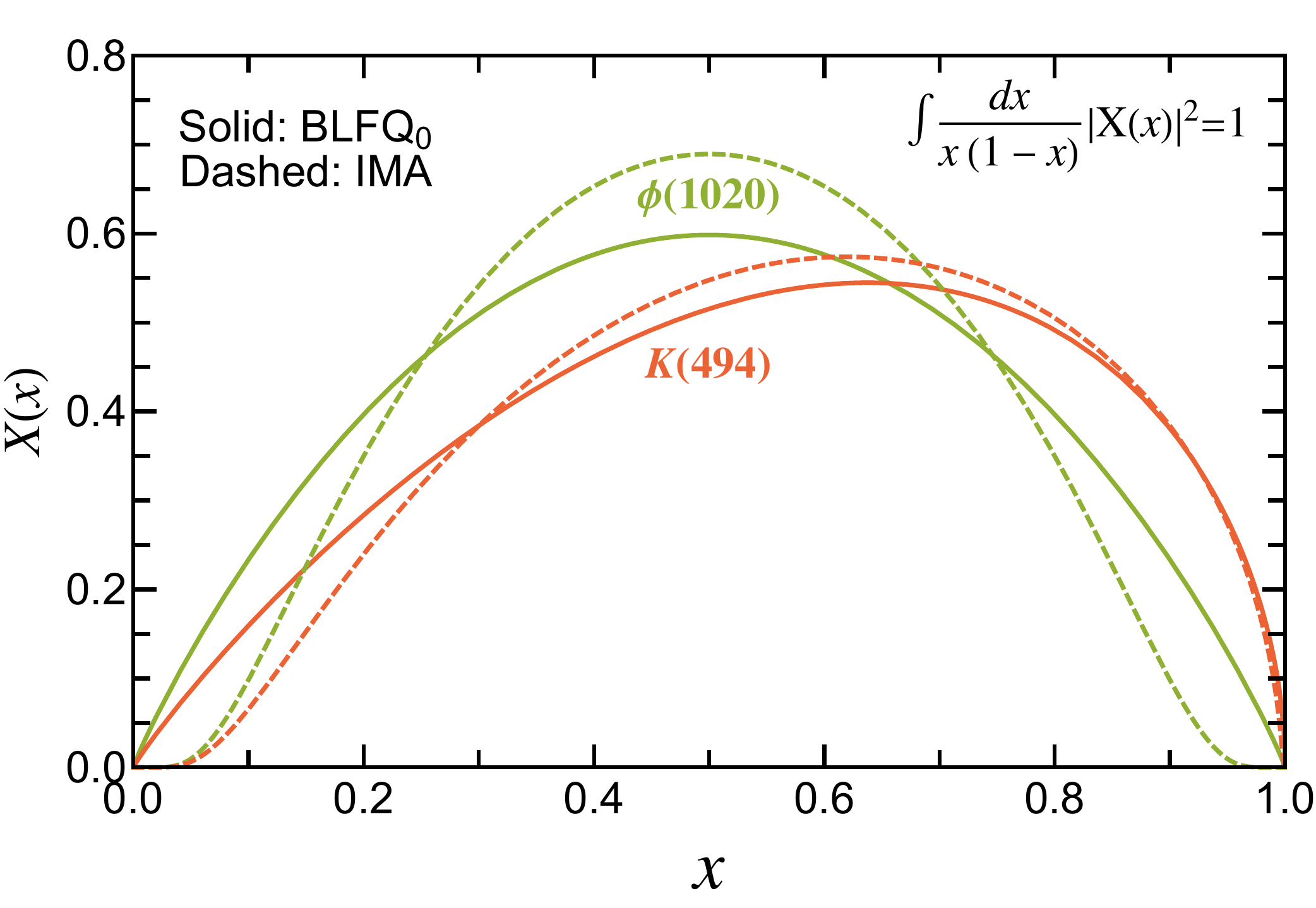}
\includegraphics[width=0.45\textwidth]{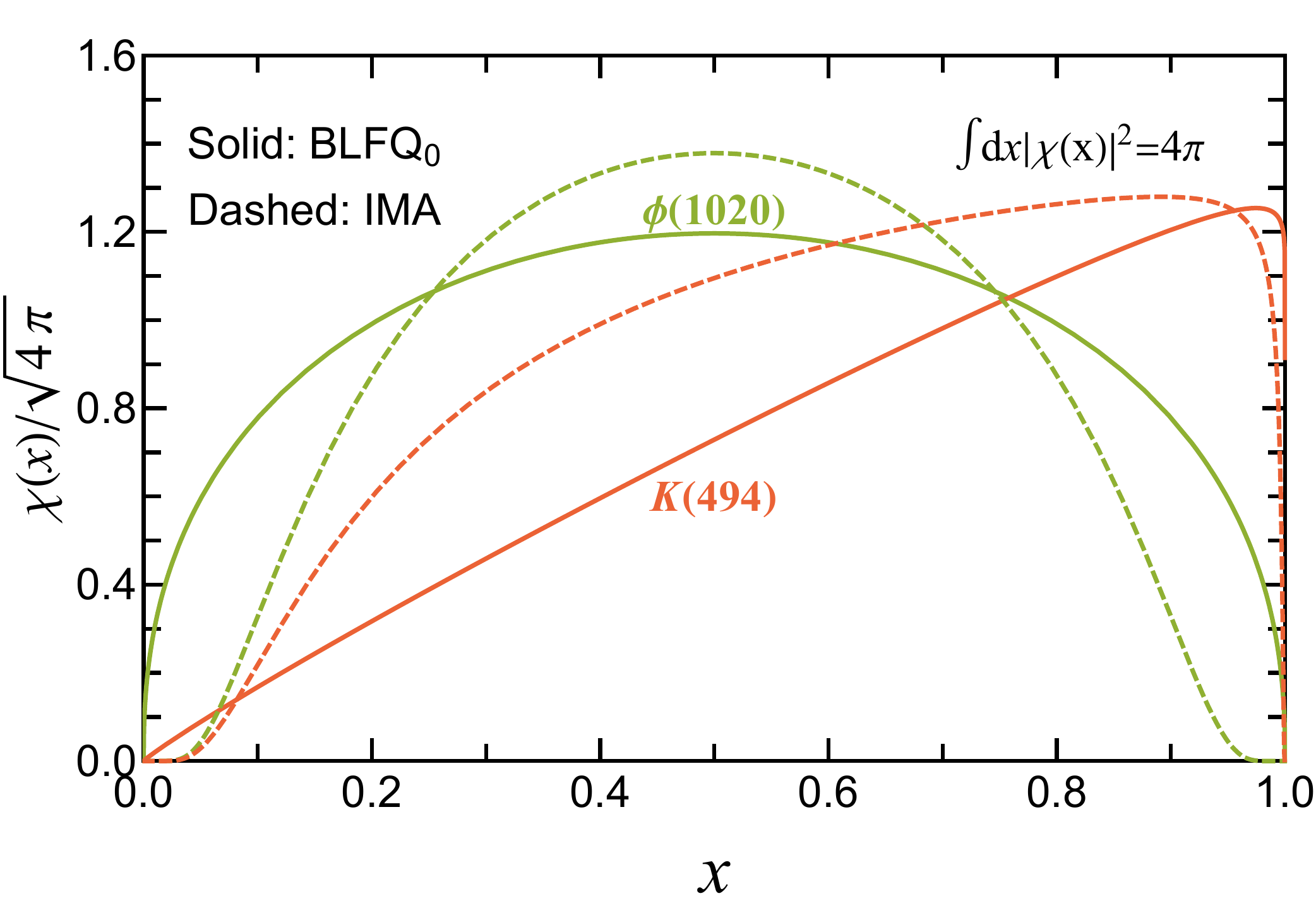}
\caption{Comparison of the longitudinal wave functions for selected light mesons obtained from the invariant mass ansatz (IMA) and from this work. The wave functions $X(x) = \sqrt{x(1-x)/4\pi}\chi(x)$ are normalized according to $\int_0^1\frac{\dd x}{x(1-x)}\big|X(x)\big|^2=1$, which is consistent with the original normalization convention of the IMA wave functions from Ref.~\cite{Brodsky:2014yha}. The parameters of this work are listed in Table~\ref{tab:model_parameters}. Note that in the IMA, the longitudinal wave functions for $\pi(140)$ and $a_0(980)$ are the same. A minus sign is applied to the wave function of $a_0(980)$ for the purpose of comparison.
}
\label{fig:IMA_vs_LCA}
\end{figure*}

Among various recent proposals, a natural candidate of the longitudinal confining interaction is the collinear QCD \cite{Burkardt:1997de} in particular the 't Hooft model \cite{tHooft:1974pnl}. Indeed, the 't Hooft model exhibits chiral symmetry breaking. Alas, the model is not analytically solvable. We instead propose an analytically solvable model for the longitudinal confining interaction that is closely related to the 't Hooft model, $V_\| = -\sigma^2 \partial_x\big( x(1-x) \partial_x \big)$, where $\sigma$ is the strength of the confinement \cite{Li:2015zda}. $\partial_x = (\partial/\partial x)|_{\vec\zeta_\perp}$ can be identified with a recently introduced frame-independent coordinate $\tilde z = (1/2)P^+x^-$ \cite{Miller:2019ysh}. 
Comparing to other form of longitudinal confinement, notably Refs.~\cite{Chabysheva:2012fe, Glazek:2013jba}, our proposal possesses several salient features \cite{Li:2015zda}. The solutions from this potential resemble the desired asymptotic parton distributions $\sim x^a(1-x)^b$ at the endpoints, which produces the GMOR relation in the vicinity of the chiral limit. In the heavy sector, the same interaction is shown to restore the 3D rotation symmetry. 
Furthermore, with this interaction, the longitudinal Schr\"odinger equation (\ref{eqn:lfse_long}) is analytically solvable \cite{Li:2015zda}. 
For brevity, we will refer to this Hamiltonian as $\mathrm{BLFQ_0}$. The resulting full mass eigenvalues are, 
\begin{multline}
M^2 = 2\kappa^2(2n+|m|+J) + \sigma (m_q+m_{\bar q})(2l+1) \\
+ \sigma^2 l(l+1) + (m_q+m_{\bar q})^2,
\end{multline}
 where $l$ is the quantum number in the longitudinal direction.
The associated longitudinal wave functions are, 
\begin{equation}
\chi_l(x) = N x^\frac{\beta}{2}(1-x)^{\frac{\alpha}{2}} P_l^{(\alpha,\beta)}(2x-1),
\end{equation}
where $N$ is a normalization constant. $P_l^{(\alpha,\beta)}(z)$ is the Jacobi polynomial and $\alpha = 2m_{\bar q}/\sigma$, $\beta = 2m_{q}/\sigma$. 
The obtained solutions are closely related to the 't~Hooft model \cite{tHooft:1974pnl, Chabysheva:2012fe} {as well as the collinear QCD in 1+1D \cite{Burkardt:1997de}}. Immediately, the mass squared of the ground state ($n=m=l=0,J=0$), i.e.~the pion, is $M_\pi^2 = 2\sigma m_{u,d} + 4m_{u,d}^2$, satisfying the GMOR relation with $\sigma = |\langle\bar q q\rangle |/f^2_\pi$, up to $\mathrm{O}(m_q)$ \cite{Colangelo:2001sp}. The pion wave function now becomes, 
$\psi_\pi(x, \vec k_\perp) = ({4\pi N}/{\kappa}) x^\frac{\mu}{2}(1-x)^\frac{\mu}{2}\exp\big[-{\vec k^2_\perp}/({2\kappa^2x(1-x)})\big]$,
where $\mu = 2m_{u,d}/\sigma$. 
Fig.~\ref{fig:IMA_vs_LCA} compares the normalized ground-state longitudinal wave functions obtained from IMA and from this work in three light meson sectors, $q\bar q$, $s\bar s$ and $s\bar q$, where $q=u,d$. Visually, the longitudinal wave functions from these two approaches are nearly identical for the $q\bar q$ sector, but differ significantly for the $s\bar s$ and $s\bar q$ sectors. Note that the quark masses from these two approaches are remarkably different (see Table \ref{tab:model_parameters} and the relevant discussions in the main text). Major differences occur at the endpoints $x\to0,1$, our wave functions are power-law like whereas the IMA wave functions are more suppressed. As a result, observables singular at the endpoints, e.g. the light-front kinetic energy and the form factors, acquire very different expectation values from these two sets of wave functions. The endpoint behavior also has a dramatic impact on hadronic observables in high energy collisions as hard kernels $T_H$ are sensitive to the endpoint singularities \cite{Lepage:1980fj}.

The incorporation of longitudinal dynamics also allows us to identify states directly with the {exact} discrete symmetries, the charge conjugation $C$ and the mirror parity $m_P=(-1)^JP$. The latter is more convenient to use than parity $P$ in light-front dynamics \cite{Brodsky:2006ez}. The parities associated with the quantum numbers $(n,m,l,S,J)$ are $m_P = (-1)^{m+S+1}$, $C=(-1)^{m+l+S}$, where $S$ is the total spin. By contrast, LFH with IMA can only describe states with $P = (-1)^{m+1}$ and $C = (-1)^{m+S}$. In general, for states without longitudinal excitations ($l=0$), e.g.~$\rho$, $\pi$, $K$, the rest of our quantum number assignments remain the same as LFH with IMA. However, when longitudinal excitations are present in the states, e.g.~$a_0$ ($0^{++}$), $b_1$ ($1^{+-}$), $a_2$ ($2^{++}$), $\rho(1700)$ ($1^{--}$), $K_1$ ($1^+$), $K_2$ ($2^-$), our quantum number assignments become different. It can be shown that our assignments are consistent with the discrete symmetries as well as with those of the traditional quark models \cite{rpp:2019}. 

Our model parameters are listed in Table~\ref{tab:model_parameters}. In LFHQCD, $\kappa$, the strength of the holographic confining potential, is obtained from the $\rho$-$\pi$ mass splitting for the light vector sector, viz $M_\rho^2 - M_\pi^2 = 2\kappa^2$, 
and $M_{K*}^2 - M_K^2 = 2\kappa^2$, while for the light pseudoscalar sector, $\kappa$ is obtained from fitting to the Regge slopes \cite{Brodsky:2014yha}. 
Note that the values of $\kappa$ obtained from the $q\bar q$ and $s\bar q$ sectors coincide, indicating the universality of the holographic confinement. 
Since the determination of $\kappa$ does not involve the longitudinal d.o.f., we adopt the same values for $\kappa$ in this work.  

The quark masses $m_{\{u,d\},s}$ in this work, in general, are different from those in LFH with IMA. As mentioned above, our quark masses satisfy the GMOR relation (\ref{eqn:GMOR}), $M^2_{\pi} = 2\sigma m_{u,d} + 4 m_{u,d}^2$, $M^2_K= \sigma (m_{u,d}+m_s) + (m_{u,d}+m_s)^2$.
Another constraint comes from the mass splitting between the vector meson ($1^{--}$) and the axial-vector meson ($1^{+-}$), which is sensitive to $\sigma$: $M^2_{b_1} - M^2_{\rho} = 2 \sigma^2 + 4 \sigma m_{u,d}$, $M^2_{K_1(1400)} - M^2_{K^*(892)} = 2 \sigma^2 + 2 \sigma (m_{u,d} + m_s)$.
The resulting quark masses $m_{u,d} = 15\,\mathrm{MeV}$, $m_s = 261\,\mathrm{MeV}$ are slightly larger than the current quark mass, yet lower than those from LFH with IMA ($m_{u,d}=46 \,\mathrm{MeV}$, $m_s = 357\,\mathrm{MeV}$), and far lower than the typical constituent quark masses ($m_{u,d} \sim 350\,\mathrm{MeV}$).  
Note that the quark masses in this work are effective quark masses at low energy resolution \cite{Brodsky:2014yha}, which are in general different from the current quark masses appearing {in, e.g., Lattice QCD \cite{FlavourLatticeAveragingGroup:2019iem}}.
Note that $\sigma$ obtained from the $q\bar q$ mesons and the $s\bar q$ mesons coincide, which is consistent with the Weinberg relation \cite{Weinberg:1977hb, McNeile:2012xh}. 
{The condensate is predicted to be $\langle \bar q q \rangle = f_\pi^2\sigma = (0.26\,\mathrm{GeV})^3$. This is compared with the standard value from chiral perturbation theory $\langle \bar q q \rangle_{\overline{\mathrm{MS}}}^{\mu=2\,\mathrm{GeV}} = (0.34\,\mathrm{GeV})^3$.}

\begin{table}
\caption{Parameters of our model. The holographic confining strength $\kappa$ is adopted from LFH \cite{Brodsky:2014yha}.}
\label{tab:model_parameters}
\centering
\begin{tabular}{ccccc}
\toprule 
$m_{u,d}$ & $m_s$ & $\kappa$ $(S=0)$ & $\kappa$ $(S=1)$ & $\sigma$ \\
\colrule
15 MeV & 261 MeV & 0.59 GeV & 0.54 GeV & 0.62 GeV \\
\botrule
\end{tabular}
\end{table}

\begin{figure}
\centering 
\includegraphics[width=1\linewidth]{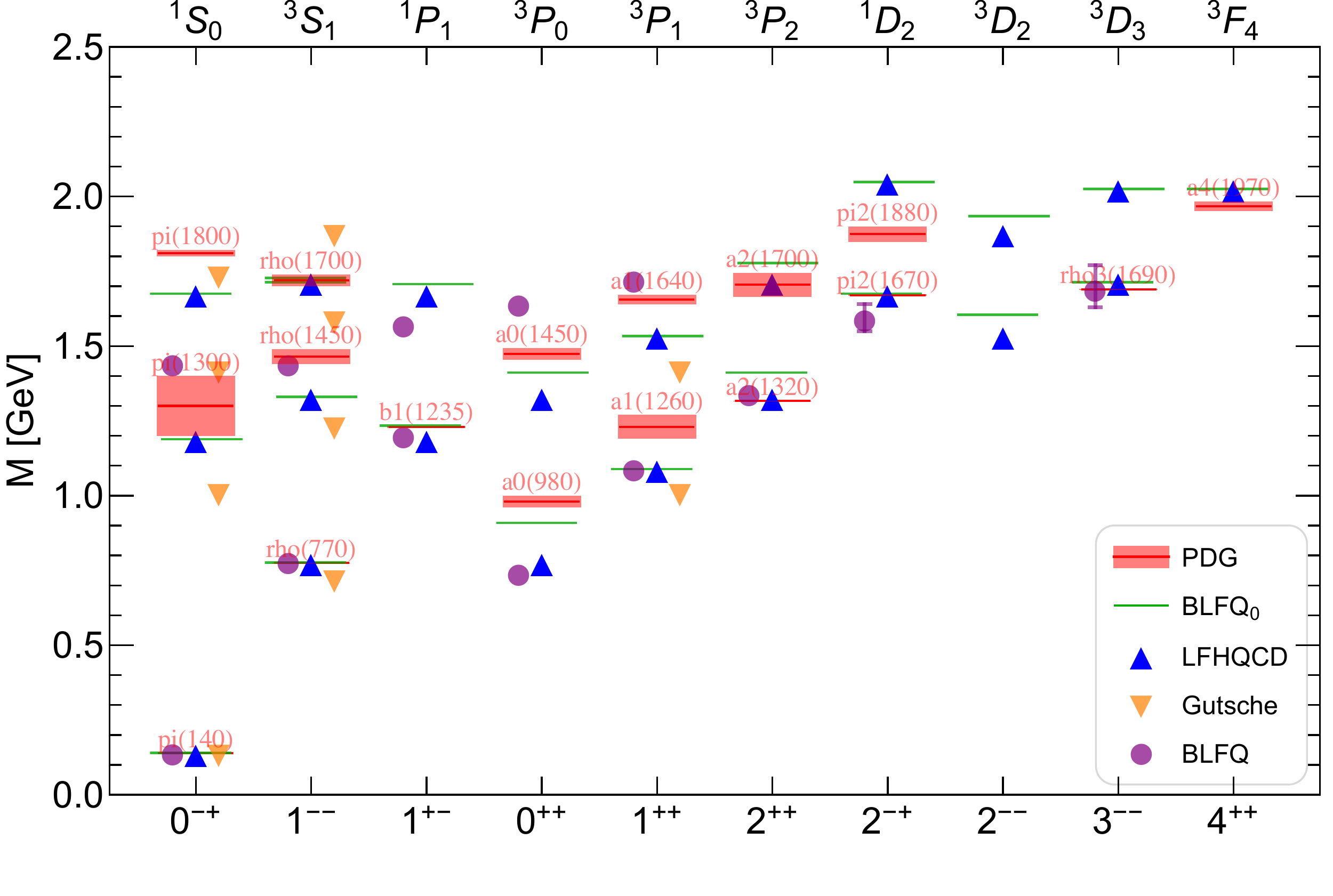} 
\includegraphics[width=1\linewidth]{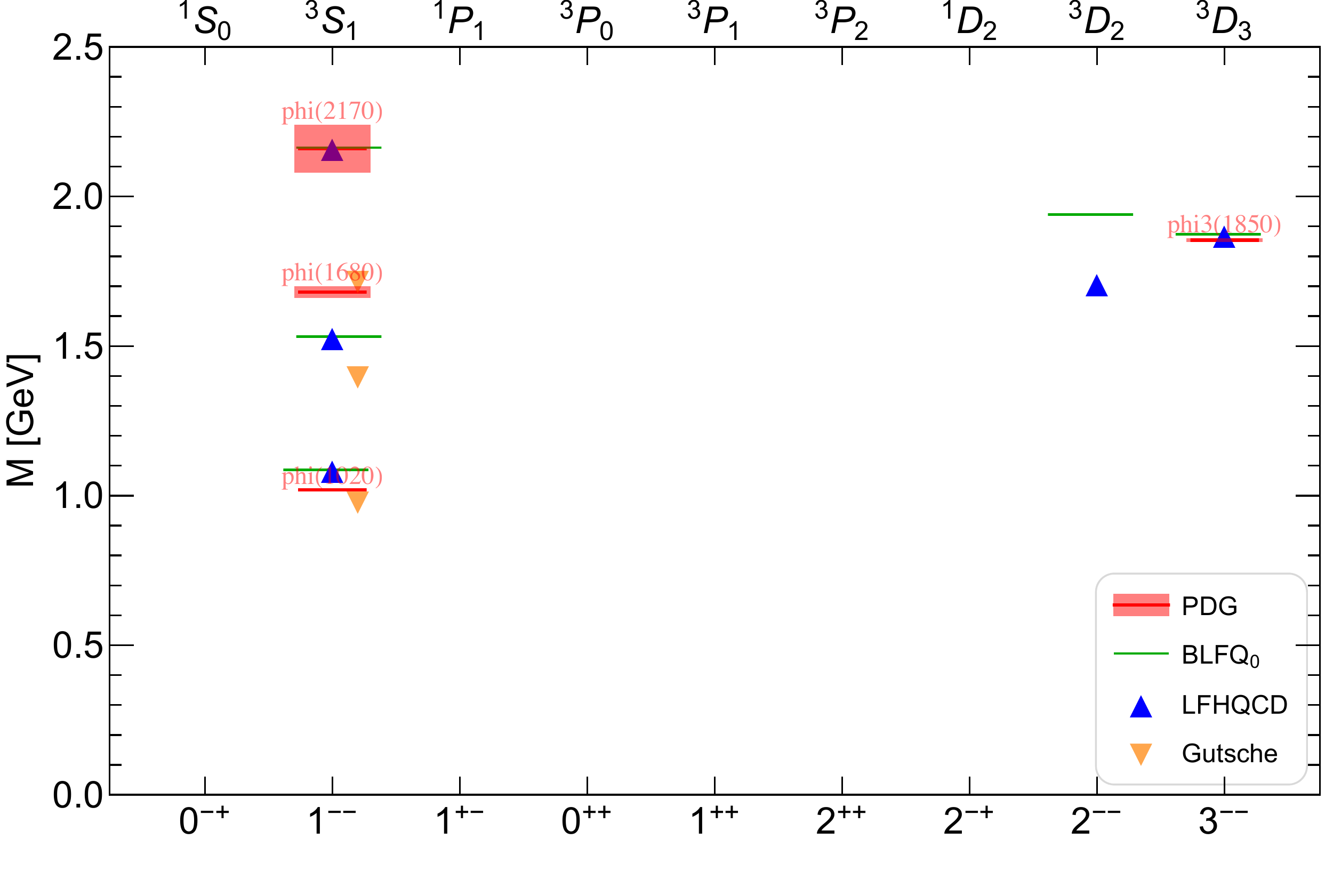} 
\includegraphics[width=1\linewidth]{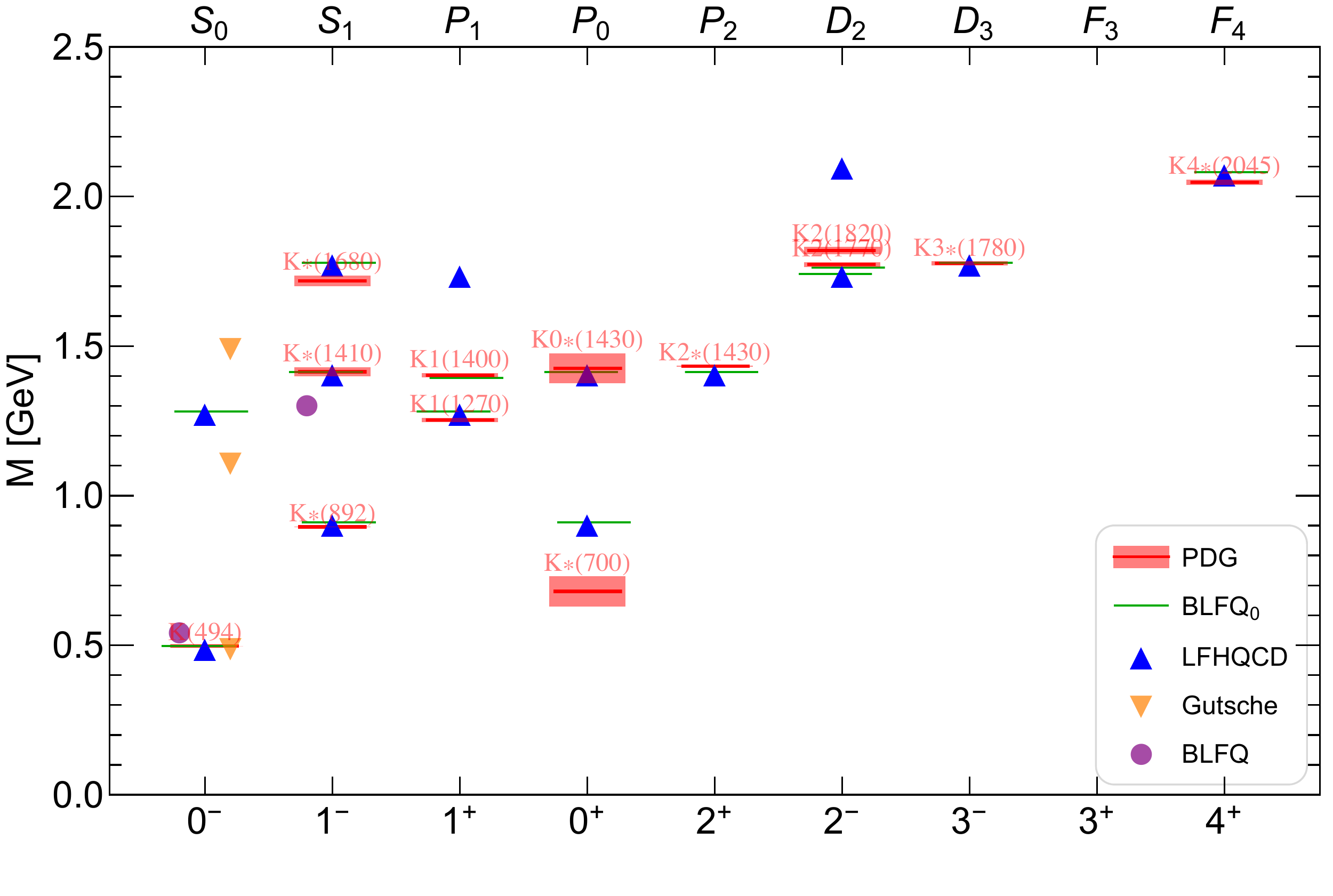}
\caption{Spectra of unflavored light mesons. The horizontal axes are $J^{\textsc{pc}}$. 
LFH and Gutsche data are taken from Refs.~\cite{Brodsky:2014yha} and \cite{Gutsche:2012ez}, respectively. 
BLFQ data are taken from Tang~et.~al.~\cite{Tang:2018myz, Tang:2019gvn} and Qian~et.~al.~\cite{Qian:2020utg}.
}
\label{fig:spectrum}
\end{figure}

Figure~\ref{fig:spectrum} shows the reconstructed light meson spectra for $q\bar q$, $s\bar s$ and $s\bar q$ systems, and compares with the experimental measurements compiled by the particle data group (PDG, \cite{pdg:2020}) and with the predictions of LFH with IMA \cite{Brodsky:2014yha}. Selected results based on LFH are also included for comparison \cite{Gutsche:2012ez, Qian:2020utg, Tang:2018myz, Tang:2019gvn}. For states without longitudinal excitation, e.g. $\pi, \rho, K, K^*, \phi$, our predictions are \textit{identical to} those of LFH with IMA. For states with longitudinal excitations, e.g. $\rho(1700), a_0, K_1, K_2$, our results improve the agreement with the experiments. 

\begin{figure*}
\centering
\includegraphics[width=0.45\linewidth]{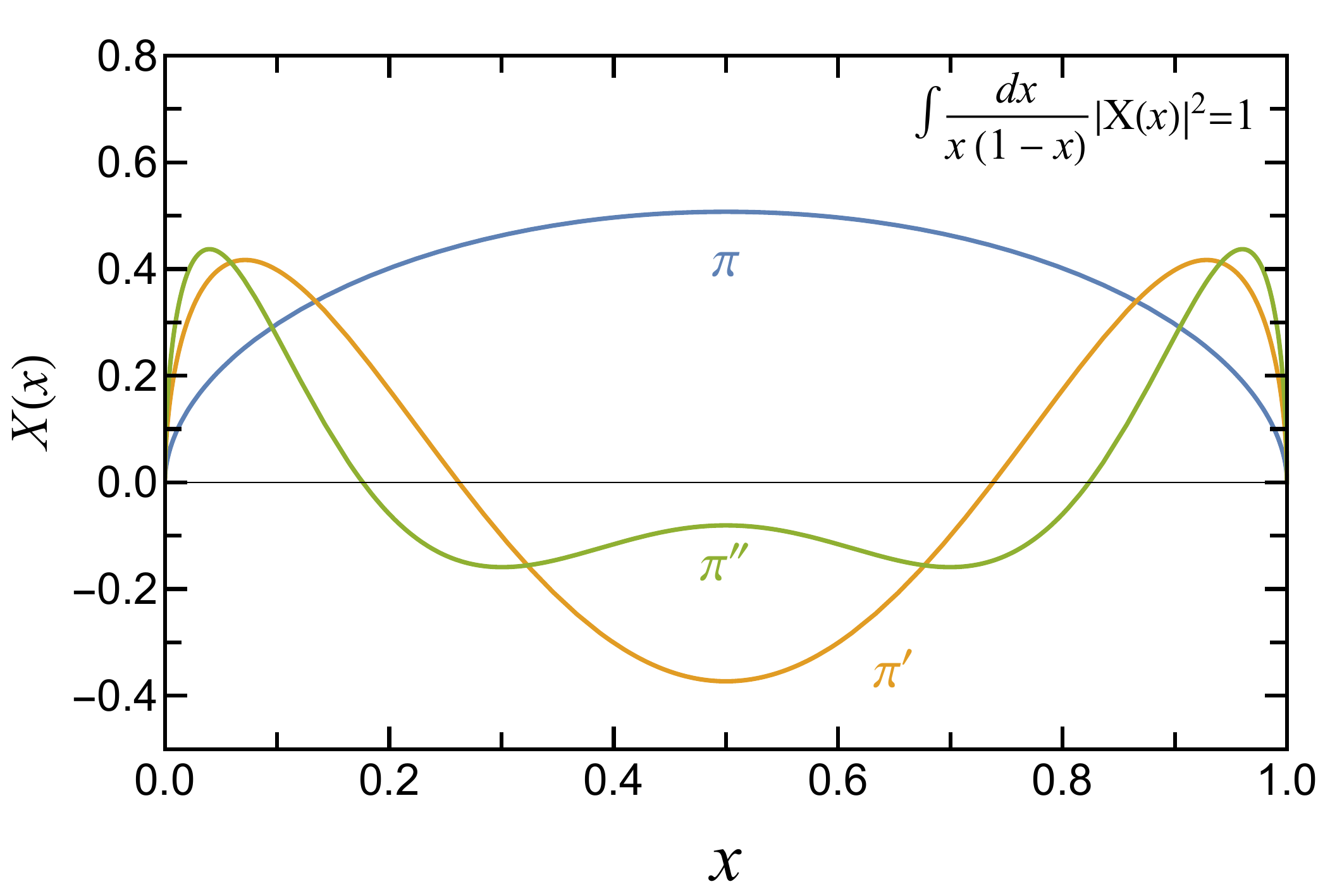}
\includegraphics[width=0.45\linewidth]{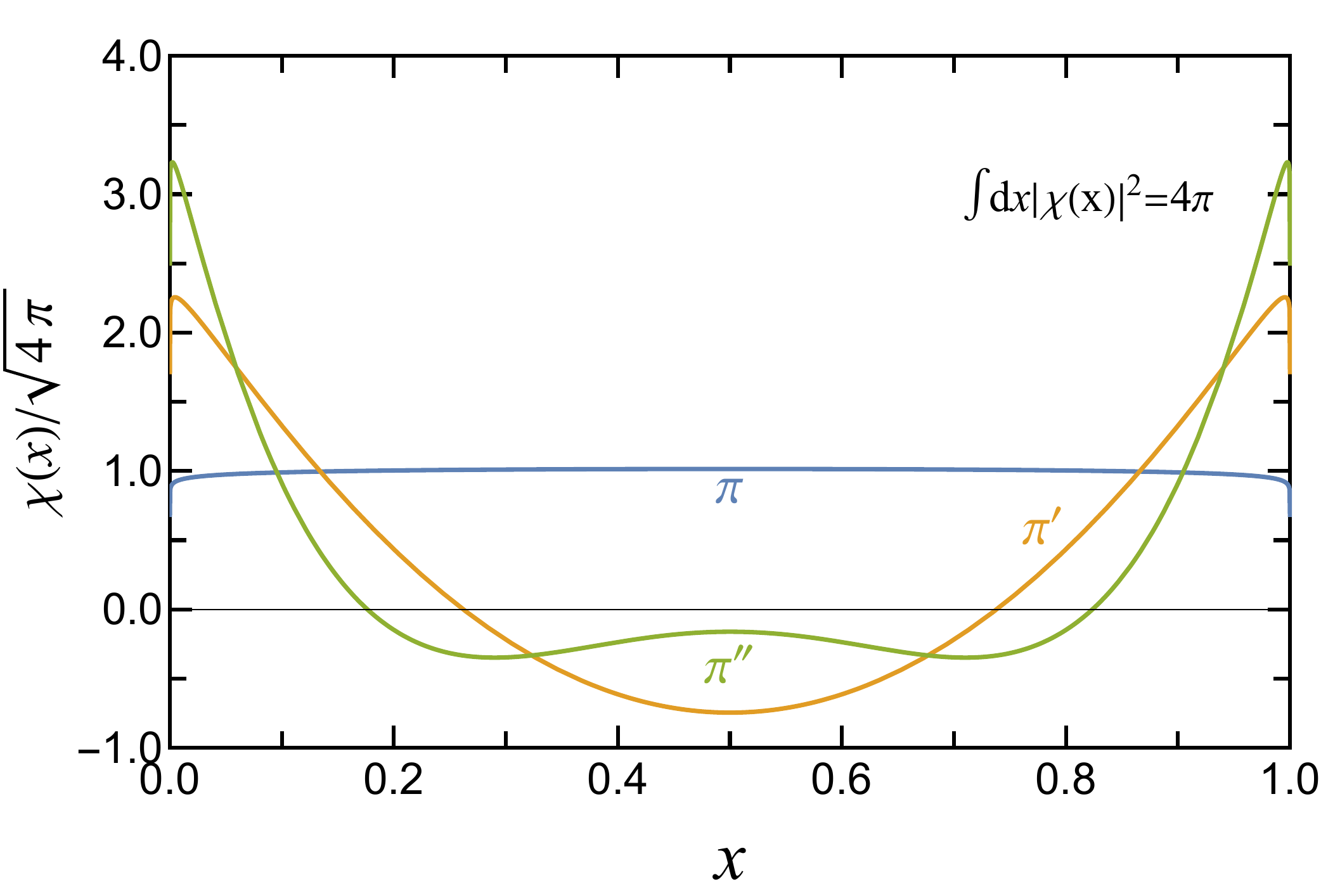}
\caption{The longitudinal wave functions of the pions. The wave function $X(x) = \sqrt{x(1-x)/4\pi}\chi(x)$ is normalized according to $\int_0^1\frac{\dd x}{x(1-x)}\big|X(x)\big|^2=1$, which is consistent with the original normalization convention of the IMA wave functions from Ref.~\cite{Brodsky:2014yha}. Under this convention, the pion distribution amplitude $\phi_\pi(x) = (\sqrt{6}\kappa/\pi)X(x)$. 
}
\label{fig:excited_pions}
\end{figure*}

The GMOR relation (\ref{eqn:GMOR}) is also applicable to excited pions. For these systems, instead of a light mass, it predicts a small decay constant.
In the chiral limit, the decay constants of the excited pions vanish exactly, indicating some highly nontrivial internal structures. In light-front dynamics, the decay constant is related to the integration of the distribution amplitude,
$f_{\pi^N} = \int_0^1 \dd x\,\phi_{\pi^N}(x)$, where the distribution amplitude,
\begin{equation}
\phi_{\pi^N}(x) = \sqrt{\frac{2N_c}{x(1-x)}} \int \frac{\dd^2 k_\perp}{(2\pi)^3} \psi_{\pi^N}(x, \vec k_\perp),
\end{equation}
 is proportional to the longitudinal wave function. As a result, to describe the radially excited pions, we also need excitations in the longitudinal direction, in addition to those in the transverse directions. We imagine that degenerate BLFQ${}_0$ states  are likely to mix significantly in future applications incorporating interactions originating in QCD \cite{Li:2017mlw}. Alternatively,
one could make phenomenological estimates of that mixing, for example, by fitting decay constants where available. Thus the physical $\pi'$ can be identified with the superposition of the $n=1, l=0$ state and the $n=0,l=2$ state. The predicted mass and decay constant are changed accordingly. 
By using the GMOR relation (\ref{eqn:GMOR}) as a constraint, we can obtain that mixing as well as the predicted mass and decay constant, as shown in Table~\ref{tab:pions}. Also shown in the table is the second excited pion $\pi''$, which is identified with the superposition of the $n=2, l=0$ state, the $n=1,l=2$ state and the $n=0,l=4$ state. The obtained decay constants of the excited pions are indeed small and the obtained masses are in rough agreement with the experimental measurements. 
Specifically, the masses $M_{\pi'}$ and $M_{\pi''}$ differ by $\sim17\%$ from the experimental values while the decay constant $f_{\pi}$ differs by $\sim 40\%$.
Fig.~\ref{fig:excited_pions} shows the resulting normalized longitudinal wave functions of the ground-state and the excited pions. 

\begin{table}
\caption{The predicted masses and decay constants of the pions obtained from the GMOR relation within our model.}
\label{tab:pions}
\begin{tabular}{cc|c c c} 
\toprule
&  & $\pi$ & $\pi'$ & $\pi''$ \\
 \colrule 
\multirow{2}{*}{This work} & $M_{\pi^N}$ & 140 MeV${}^*$ & 1.52 GeV & 2.12 GeV \\
& $f_{\pi^N}$ & 182 MeV & 12 MeV & 8.5 MeV \\
\colrule
\multirow{1.5}{*}{experiments} & $M_{\pi^N}$ & 140 MeV & 1.30 GeV & 1.81 GeV \\
\multirow{0.5}{*}{(PDG)} & $f_{\pi^N}$ & 130 MeV &   &  \\
\botrule
\end{tabular}
\end{table}

In this work, we have introduced a 3D semiclassical bound-state equation for hadrons that extends light-front holography by taking advantage of the residual degree of freedom in the longitudinal direction. By requiring power-law like asymptotics at the endpoints ($x\sim 0,1$) in the longitudinal wave functions, we arrived at a specific effective longitudinal confinement term, which produces analytically solvable mass spectra and light-front wave functions. The obtained light meson spectra are consistent with the Gell-Mann-Oakes-Renner relation as required by the chiral symmetry breaking and are in better agreement with the experiments. Compared to other longitudinal confining models, our model is particularly suitable as a starting point for future quantum many-body calculations. 

One of the major advantages of the present model is that, for states without longitudinal excitation, e.g.~$\pi$, $\rho$, $K$, which are those attracting most of the attention, the predicted masses and wave functions are very close to the predictions of LFH. Thus, leading phenomenological successes of LFH remain intact in the present approach {(cf.~\cite{Jia:2018ary})}. Exceptions may arise for observables sensitive to the endpoints. One notable example is the elastic form factor, $F_\pi(Q^2) = \int_0^1\dd x \chi^2(x) \exp \big[-((1-x){Q^2})/(4x\kappa^2)\big]$. At small $Q^2$, the pion form factor obtained from our work is close to that from LFH with IMA. However, at large $Q^2$, form factor from LFH with IMA scales as $Q^2F_\pi(Q^2) \to \exp(-c Q^2)$, whereas from our model $Q^2F_\pi(Q^2) \to 1/(Q^2)^\mu \sim 1/\ln Q^2$ since $\mu=2m_{u,d}/\sigma \ll 1$. Further work is needed to examine the connections of the longitudinal confinement with both the quark condensate (cf. in-hadron condensate \cite{Brodsky:2010xf}) as well as with the successful heavy quark effective theory. It will also be interesting to explore the formal link between the present model and the top-down holographic models, which would provide useful guidance for physics beyond the semiclassical approximation. 

Quantum corrections are absent in the present model, as indicated by the deviation of the quark masses from the current quark masses. One can systematically improve the model by incorporating more realistic effective interactions. The leading-order effective Hamiltonian derived from RGPEP has been shown to be consistent with the present model in the nonrelativistic limit \cite{Glazek:2017rwe}. 
In Ref.~\cite{Li:2017mlw}, we apply the same model to heavy quarkonium with an additional one-gluon exchange interaction derived from the Okubo-Suzuki-Lee-Wilson-Bloch renormalization procedure. The model is solved in BLFQ and the mass spectra and properties of heavy quarkonia are well reproduced. This is the first step towards a full relativistic quantum many-body treatment of light-front QCD. 

 {We wish to thank S.J. Brodsky and G.F. de T\'eramond for enlightening discussions and for their critical reading of the manuscript. We also gratefully acknowledge fruitful discussions with P. Mannheim, X. Zhao, P. Maris, G. Miller, C. Mondal, L. Zhang, and M. Huang.} This work is supported in part by the Department of Energy under Grants No. DE-FG02-87ER40371, and No. DE-SC0018223 (SciDAC4/NUCLEI). 

\textit{Note added:} After the appearance of our work, similar work addressing the longitudinal dynamics appeared. de Téramond and Brodsky arXiv:2103.10950 [hep-ph] adopted the same longitudinal interaction and focused on the ground states including the heavy quarkonia. Ahmady et al. [Phys. Rev. D 104, 7 (2021) \& Phys. Lett. B \textbf{823}, 136754 (2021)] adopted the 't Hooft interaction as longitudinal confining potential, as was first suggested by Chabysheva and Hiller \cite{Chabysheva:2012fe}. A phenomenological comparison of the hadron spectra employing the 't Hooft model and our model is also reported. Finally, these longitudinal confining potentials and the corresponding wave functions also critically compared by Weller \& Miller in arXiv:2111.03194 [hep-ph].


\begin{thebibliography}{99}

\bibitem{Vary:2009gt}
J.~P.~Vary, H.~Honkanen, J.~Li, P.~Maris, S.~J.~Brodsky, A.~Harindranath, G.~F.~de Teramond, P.~Sternberg, E.~G.~Ng and C.~Yang,
``Hamiltonian light-front field theory in a basis function approach,''
Phys. Rev. C \textbf{81}, 035205 (2010);
[arXiv:0905.1411 [nucl-th]].


\bibitem{Glazek:2012qj}
S.~D.~Glazek,
``Perturbative formulae for relativistic interactions of effective particles,''
Acta Phys. Polon. B \textbf{43}, 1843-1862 (2012);
[arXiv:1204.4760 [hep-th]].


\bibitem{Glazek:2017rwe}
S.~D.~G\l{}azek, M.~G\'omez-Rocha, J.~More and K.~Serafin,
``Renormalized quark\textendash{}antiquark Hamiltonian induced by a gluon mass ansatz in heavy-flavor QCD,''
Phys. Lett. B \textbf{773}, 172-178 (2017);
[arXiv:1705.07629 [hep-ph]].

\bibitem{Karch:2006pv}
A.~Karch, E.~Katz, D.~T.~Son and M.~A.~Stephanov,
``Linear confinement and AdS/QCD,''
Phys. Rev. D \textbf{74}, 015005 (2006);
[arXiv:hep-ph/0602229 [hep-ph]].


\bibitem{Brodsky:2013ar}
S.~J.~Brodsky, G.~F.~De T\'eramond and H.~G.~Dosch,
``Threefold Complementary Approach to Holographic QCD,''
Phys. Lett. B \textbf{729}, 3-8 (2014);
[arXiv:1302.4105 [hep-th]].

\bibitem{deTeramond:2014asa}
G.~F.~de Teramond, H.~G.~Dosch and S.~J.~Brodsky,
``Baryon Spectrum from Superconformal Quantum Mechanics and its Light-Front Holographic Embedding,''
Phys. Rev. D \textbf{91}, no.4, 045040 (2015);
[arXiv:1411.5243 [hep-ph]].

\bibitem{Brodsky:2006uqa}
S.~J.~Brodsky and G.~F.~de Teramond,
``Hadronic spectra and light-front wavefunctions in holographic QCD,''
Phys. Rev. Lett. \textbf{96}, 201601 (2006);
[arXiv:hep-ph/0602252 [hep-ph]].

\bibitem{Dosch:2016zdv}
H.~G.~Dosch, G.~F.~de Teramond and S.~J.~Brodsky,
``Supersymmetry Across the Light and Heavy-Light Hadronic Spectrum II,''
Phys. Rev. D \textbf{95}, no.3, 034016 (2017);
[arXiv:1612.02370 [hep-ph]].

%
\bibitem{Dosch:2015bca}
H.~G.~Dosch, G.~F.~de Teramond and S.~J.~Brodsky,
``Supersymmetry Across the Light and Heavy-Light Hadronic Spectrum,''
Phys. Rev. D \textbf{92}, no.7, 074010 (2015);
[arXiv:1504.05112 [hep-ph]].


\bibitem{Nielsen:2018ytt}
M.~Nielsen, S.~J.~Brodsky, G.~F.~de T\'eramond, H.~G.~Dosch, F.~S.~Navarra and L.~Zou,
``Supersymmetry in the Double-Heavy Hadronic Spectrum,''
Phys. Rev. D \textbf{98}, no.3, 034002 (2018);
[arXiv:1805.11567 [hep-ph]].

\bibitem{Zou:2019tpo}
L.~Zou, H.~G.~Dosch, G.~F.~De T\'eramond and S.~J.~Brodsky,
``Isoscalar mesons and exotic states in light front holographic QCD,''
Phys. Rev. D \textbf{99}, no.11, 114024 (2019);
[arXiv:1901.11205 [hep-ph]].


\bibitem{Brodsky:2008pf}
S.~J.~Brodsky and G.~F.~de Teramond,
``Light-Front Dynamics and AdS/QCD Correspondence: Gravitational Form Factors of Composite Hadrons,''
Phys. Rev. D \textbf{78}, 025032 (2008);
[arXiv:0804.0452 [hep-ph]].

\bibitem{Sufian:2016hwn}
R.~S.~Sufian, G.~F.~de T\'eramond, S.~J.~Brodsky, A.~Deur and H.~G.~Dosch,
``Analysis of nucleon electromagnetic form factors from light-front holographic QCD : The spacelike region,''
Phys. Rev. D \textbf{95}, no.1, 014011 (2017);
[arXiv:1609.06688 [hep-ph]].

\bibitem{deTeramond:2018ecg}
G.~F.~de Teramond \textit{et al.} [HLFHS],
``Universality of Generalized Parton Distributions in Light-Front Holographic QCD,''
Phys. Rev. Lett. \textbf{120}, no.18, 182001 (2018);
[arXiv:1801.09154 [hep-ph]].

\bibitem{Liu:2019vsn}
T.~Liu, R.~S.~Sufian, G.~F.~de T\'eramond, H.~G.~Dosch, S.~J.~Brodsky and A.~Deur,
``Unified Description of Polarized and Unpolarized Quark Distributions in the Proton,''
Phys. Rev. Lett. \textbf{124}, no.8, 082003 (2020);
[arXiv:1909.13818 [hep-ph]].

 \bibitem{Brodsky:2014yha} 
  S.~J.~Brodsky, G.~F.~de Teramond, H.~G.~Dosch and J.~Erlich,
  ``Light-Front Holographic QCD and Emerging Confinement,''
  Phys.\ Rept.\  {\bf 584}, 1 (2015);
  [arXiv:1407.8131 [hep-ph]].
  
\bibitem{Brodsky:2020ajy}
S.~J.~Brodsky, G.~F.~de Teramond and H.~G.~Dosch,
``Light-Front Holography and Supersymmetric Conformal Algebra: A Novel Approach to Hadron Spectroscopy, Structure, and Dynamics,''
[arXiv:2004.07756 [hep-ph]].

\bibitem{Gell-Mann:1968}
Murray Gell-Mann, R. J. Oakes, and B. Renner, 
``Behavior of Current Divergences under $SU_3\times SU_3$, 
Phys. Rev. \textbf{175}, 2195 (1968).

\bibitem{Burkardt:1997de}
M.~Burkardt,
``Mesons in a collinear QCD model,''
Phys. Rev. D \textbf{56}, 7105-7118 (1997);
[arXiv:hep-ph/9705224 [hep-ph]].

\bibitem{tHooft:1974pnl}
G.~'t Hooft,
``A Two-Dimensional Model for Mesons,''
Nucl. Phys. B \textbf{75}, 461-470 (1974).

\bibitem{Li:2015zda} 
Y.~Li, P.~Maris, X.~Zhao and J.~P.~Vary,
``Heavy Quarkonium in a Holographic Basis,''
Phys.\ Lett.\ B {\bf 758}, 118 (2016);
[arXiv:1509.07212 [hep-ph]].
  
  \bibitem{Miller:2019ysh}
G.~A.~Miller and S.~J.~Brodsky,
``Frame-independent spatial coordinate $\tilde{z}$: Implications for light-front wave functions, deep inelastic scattering, light-front holography, and lattice QCD calculations,''
Phys. Rev. C \textbf{102}, no.2, 022201 (2020);
[arXiv:1912.08911 [hep-ph]].

\bibitem{Chabysheva:2012fe}
S.~S.~Chabysheva and J.~R.~Hiller,
``Dynamical model for longitudinal wave functions in light-front holographic QCD,''
Annals Phys. \textbf{337}, 143-152 (2013);
[arXiv:1207.7128 [hep-ph]].

\bibitem{Glazek:2013jba}
S.~D.~Glazek and A.~P.~Trawi\'nski,
``Model of the AdS/QFT duality,''
Phys. Rev. D \textbf{88}, no.10, 105025 (2013);
[arXiv:1307.2059 [hep-ph]].

\bibitem{Colangelo:2001sp}
G.~Colangelo, J.~Gasser and H.~Leutwyler,
Phys. Rev. Lett. \textbf{86}, 5008-5010 (2001)
doi:10.1103/PhysRevLett.86.5008
[arXiv:hep-ph/0103063 [hep-ph]].

\bibitem{Lepage:1980fj}   
  G.~P.~Lepage and S.~J.~Brodsky,
  Phys.\ Rev.\ D {\bf 22}, 2157 (1980).

\bibitem{Brodsky:2006ez}
S.~J.~Brodsky, S.~Gardner and D.~S.~Hwang,
``Discrete symmetries on the light front and a general relation connecting nucleon electric dipole and anomalous magnetic moments,''
Phys. Rev. D \textbf{73}, 036007 (2006);
[arXiv:hep-ph/0601037 [hep-ph]].

\bibitem{rpp:2019}
M. Tanabashi et al. (Particle Data Group), 
``2019 Review of Particle Physics''
Phys. Rev. D 98, 030001 (2018) and 2019 update.

\bibitem{FlavourLatticeAveragingGroup:2019iem}
S.~Aoki \textit{et al.} [Flavour Lattice Averaging Group],
Eur. Phys. J. C \textbf{80}, no.2, 113 (2020);
[arXiv:1902.08191 [hep-lat]].

\bibitem{Weinberg:1977hb}
S.~Weinberg,
``The Problem of Mass,''
Trans. New York Acad. Sci. \textbf{38}, 185-201 (1977);

\bibitem{McNeile:2012xh}
C.~McNeile, A.~Bazavov, C.~T.~H.~Davies, R.~J.~Dowdall, K.~Hornbostel, G.~P.~Lepage and H.~D.~Trottier,
``Direct determination of the strange and light quark condensates from full lattice QCD,''
Phys. Rev. D \textbf{87}, no.3, 034503 (2013);
[arXiv:1211.6577 [hep-lat]].

\bibitem{pdg:2020}
P.A. Zyla et al. (Particle Data Group), Prog. Theor. Exp. Phys. 2020, \textbf{083C0}1 (2020).

  \bibitem{Gutsche:2012ez} 
  T.~Gutsche, V.~E.~Lyubovitskij, I.~Schmidt and A.~Vega,
  ``Chiral Symmetry Breaking and Meson Wave Functions in Soft-Wall AdS/QCD,''
  Phys.\ Rev.\ D {\bf 87}, no. 5, 056001 (2013);
  [arXiv:1212.5196 [hep-ph]].

\bibitem{Tang:2018myz}
S.~Tang, Y.~Li, P.~Maris and J.~P.~Vary,
``$B_c$ mesons and their properties on the light front'',
Phys. Rev. D \textbf{98}, no.11, 114038 (2018);
[arXiv:1810.05971 [nucl-th]].

\bibitem{Tang:2019gvn}
S.~Tang, Y.~Li, P.~Maris and J.~P.~Vary,
``Heavy-light mesons on the light front'',
Eur. Phys. J. C \textbf{80}, no.6, 522 (2020);
[arXiv:1912.02088 [nucl-th]].

\bibitem{Qian:2020utg}
W.~Qian, S.~Jia, Y.~Li and J.~P.~Vary,
``Light mesons within the basis light-front quantization framework,''
Phys. Rev. C \textbf{102}, no.5, 055207 (2020);
[arXiv:2005.13806 [nucl-th]].

\bibitem{Jia:2018ary}
S.~Jia and J.~P.~Vary,
``Basis light front quantization for the charged light mesons with color singlet Nambu\textendash{}Jona-Lasinio interactions,''
Phys. Rev. C \textbf{99}, no.3, 035206 (2019);
[arXiv:1811.08512 [nucl-th]].

\bibitem{Brodsky:2010xf}
S.~J.~Brodsky, C.~D.~Roberts, R.~Shrock and P.~C.~Tandy,
``Essence of the vacuum quark condensate,''
Phys. Rev. C \textbf{82}, 022201 (2010);
[arXiv:1005.4610 [nucl-th]].

\bibitem{Li:2017mlw}
Y.~Li, P.~Maris and J.~P.~Vary,
``Quarkonium as a relativistic bound state on the light front,''
Phys. Rev. D \textbf{96}, no.1, 016022 (2017);
[arXiv:1704.06968 [hep-ph]].



\end{thebibliography}
\end{document}